\documentclass[superscriptaddress,pra]{revtex4}

\usepackage[latin1]{inputenc}
\usepackage{graphicx}
\usepackage{amsmath}
\usepackage{amssymb}
\usepackage{dsfont}
\usepackage{graphicx}

\makeatletter
\def\erf{\mathop{\operator@font erf}\nolimits}
\makeatother

\newcommand\be{\begin{equation}}
\newcommand\ee{\end{equation}}

\begin{document}

\title{Unifying distribution functions: some lesser known distributions}
\author{J.R. Moya-Cessa,$^{1}$ ,H. Moya-Cessa,$^{2*}$ L.R. Berriel-Valdos,$^{2}$ O. Aguilar-Loreto,$^{2}$ and
P. Barberis-Blostein$^{2}$} \affiliation{$^{1}$Centro de
Investigaciones en Optica, A.C., Apartado Postal 1-948, Le\'on,
Gto., Mexico, \\$^{2}$INAOE, Coordinaci\'on de Optica, Apdo.
Postal 51 y 216, 72000 Puebla, Pue., Mexico,}

\begin{abstract}
We show that there is a way to unify distribution functions that
describe simultaneously a signal in space and (spatial) frequency.
Probably the most known of them is the Wigner distribution
function. Here we show how to unify functions of the Cohen class,
Rihacek's complex energy function, Husimi and Glauber-Sudarshan
distribution functions.
\end{abstract}
 \maketitle

\section{Introduction}\label{intro}
Distribution functions are widely used in optical physics (see
\cite{Baastians} for a review) and in quantum mechanics where they
are usually called quasiprobability distribution functions
\cite{Hillery,Schleich,Moya}). Probably the best known
quasiprobability distribution  is the Wigner function
\cite{Hillery,Wigner32} with applications in reconstruction of
signals \cite{Wolf} and resolution \cite{Berriel} in the classical
world and reconstruction  of quantum states of different systems
such as ions \cite{Wineland} or quantized fields
\cite{Moya,Haroche,Moya2,Moya3} in the quantum world. In this
contribution we would like to re-introduce a lesser known
quasiprobability distribution function, namely the
Kirkwood-Rihaczek funtion
\cite{Baastians,Kirkwood,Rihaczek,Praxmayer1,Praxmayer2}, show how
it can be related to the Wigner function, and express it as an
expectation value in some eigenbasis, just as the other
quasiprobability functions may be also expressed.

\section{Best known quasiprobability distribution functions}
\subsection{Wigner function}
We start by introducing the Wigner function, probably the best
known. It may be written in two forms: series representation (see
for instance \cite{series}), and integral representation
\begin{equation}
W(q,p) = \frac{1}{2\pi}\int du e^{iup}\langle q+\frac{u}{2}|\rho
|q-\frac{u}{2}\rangle , \label{wigner}
\end{equation}
for simplicity we use the Dirac notation here (see the appendix).
In the above equation, $\rho$ is the so-called density matrix. In
1932, Wigner introduced this function $W(q,p)$, known now as
 his distribution function \cite{Wigner32,Hillery} and contains
complete information about the state of the system ($\rho=\vert
\psi \rangle \langle\psi\vert$).

It may be written also as in terms of the (double) Fourier
transform of the characteristic function
\begin{equation}
W(\alpha )= \frac{1}{4\pi ^{2}}\int \exp (\alpha \beta ^{\ast
}-\alpha ^{\ast }\beta )C(\beta )d^{2}\beta, \label{wigcar}
\end{equation}
with $\alpha=(q+ip)/\sqrt{2}$ and where $C(\beta )$ in terms of
annihilation and creation operators is given by
\begin{equation}
C(\beta )=Tr\{\hat{\rho}\exp (\beta \hat{a}^{\dagger }-\beta ^{\ast }\hat{a}%
)\}, \label{caracteristic}
\end{equation}
also known as ambiguity function in classical optics
\cite{Stenholm}.

\subsection{$Q$-function}

The $Q$  or Husimi function \cite{Husimi}, which is expressed as
the coherent state expectation value of the density operator

\begin{equation}
Q(\alpha )=\ \frac{1}{\pi ^{2}}\int \exp (\alpha \beta ^{\ast
}-\alpha
^{\ast }\beta )Tr\{\hat{\rho}\exp (-\beta ^{\ast }\hat{a})\exp (\beta \hat{a}%
^{\dagger })\}d^{2}\beta , \label{anti}
\end{equation}
the alternative form is

\begin{equation}
Q(\alpha )=\frac{1}{\pi }\left\langle \alpha \left\vert \hat{\rho}%
\right\vert \alpha \right\rangle,
\end{equation}

\subsection{Relating $Q$ and Wigner functions in a differential form}

It is possible two group the Wigner and  the Husimi functions

\begin{equation}
F(\alpha ,s)=\frac{1}{\pi }\int C(\beta ,s)\exp (\alpha \beta
^{\ast }-\alpha ^{\ast }\beta )d^{2}\beta
\end{equation}
where $C(\beta ,s)$ is the characteristic function of order $s$

\begin{equation}
C(\beta ,s)=Tr\{\hat{D}(\beta )\hat{\rho}\}\exp (s\left\vert \beta
\right\vert ^{2}/2)
\end{equation}
with  $s$ a parameter that defines which is the function we are
looking at. For $s=1$ it is obtained the $P$-function, for $s=0$
the Wigner function, and for $s=-1$  the $Q$-function.

The $Q$-function is then

\begin{equation}
Q(\alpha )=\int G(\beta )\exp (\alpha \beta ^{\ast }-\alpha ^{\ast
}\beta )d^{2}\beta
\end{equation}
and for  $s=0$ the  Wigner  function\bigskip
\begin{equation}
W(\alpha )=\int G(\beta )\exp (\alpha \beta ^{\ast }-\alpha ^{\ast
}\beta )\exp (\left\vert \beta \right\vert ^{2}/2)d^{2}\beta
\label{ser}
\end{equation}
where
\begin{equation} G(\beta )=\frac{1}{\pi ^{2}}Tr\{D(\beta
)\hat{\rho}\}\exp (-\left\vert \beta \right\vert ^{2}/2),
\end{equation}
The equation above may be written as an  infinite (Taylor) series
 and inserted into (\ref{ser}) to obtain
\begin{equation}
W(\alpha )=\sum\limits_{n=0}^{\infty }\frac{2^{-n}}{n!}\int
G(\beta )\exp (\alpha \beta ^{\ast }-\alpha ^{\ast }\beta
)\left\vert \beta \right\vert ^{2n}d^{2}\beta. \label{seri}
\end{equation}
Considering the equality

\begin{equation}
\frac{\partial }{\partial \alpha }\frac{\partial }{\partial \alpha ^{\ast }}%
\exp (\alpha \beta ^{\ast }-\alpha ^{\ast }\beta )\bigskip
=-\left\vert \beta \right\vert ^{2}\exp (\alpha \beta ^{\ast
}-\alpha ^{\ast }\beta )
\end{equation}
we can cast equation (\ref{seri}) into
\begin{equation}
W(\alpha )=\sum\limits_{n=0}^{\infty }\frac{2^{-n}}{n!}\left( -\frac{%
\partial }{\partial \alpha }\frac{\partial }{\partial \alpha ^{\ast }}%
\right) ^{n}Q(\alpha ),
\end{equation}
or, finally \cite{series}
\begin{equation}
W(\alpha )=\exp \left( -\frac{1}{2}\frac{\partial }{\partial \alpha }\frac{%
\partial }{\partial \alpha ^{\ast }}\right) Q(\alpha ).
\end{equation}
The analysis just done will help us to relate the
Kirkwood-Rihaczek function with the Wigner function.
 \subsection{Glauber-Sudarshan}
For the sake of completeness, we introduce another well-known
distribution function: the Glauber-Sudarshan $P$-function
\cite{Glauber,Sudarshan}. First let us note that by using the
coherent state eigenbasis, we can express the density matrix as
the following double integral

\begin{equation}
\hat{\rho}=\frac{1}{\pi ^{2}}\int \int \left\langle \alpha \right\vert \hat{%
\rho}\left\vert \beta \right\rangle \left\vert \alpha
\right\rangle \left\langle \beta \right\vert d^{2}\alpha
d^{2}\beta.
\end{equation}
This representation involves off-diagonal elements  $ \left\langle
\alpha \right\vert \hat{\rho}\left\vert \beta \right\rangle ,$ and
two integrations in phase space. The next diagonal representation
was introduced independently by Glauber and Sudarshan
\cite{Glauber,Sudarshan}
\begin{equation}
\hat{\rho}=\int P(\alpha )\left\vert \alpha \right\rangle
\left\langle \alpha \right\vert d^{2}\alpha \label{P-function}
\end{equation}
and involves only one integration. Using the equation above, we
can write the Glauber-Sudarshan $P$-function in the form
 \begin{equation} P(\alpha )=\
\frac{1}{\pi ^{2}}\int \exp (\alpha \beta ^{\ast }-\alpha
^{\ast }\beta )Tr\{\hat{\rho}\exp (\beta \hat{a}%
^{\dagger })\exp (-\beta ^{\ast }\hat{a})\}d^{2}\beta,
\label{normal}
\end{equation}
or
\begin{equation}
P(\alpha )=F(\alpha ,1)=\frac{1}{\pi }\int C(\beta ,1)\exp (\alpha
\beta ^{\ast }-\alpha ^{\ast }\beta )d^{2}\beta
\end{equation}
\subsection{Cohen-class distribution functions}
A function of the Cohen class is described by the general formula
\cite{Baastians,Cohen}
\begin{equation}
W_C=\frac{1}{2\pi}\int\int\int\phi(y+\frac{1}{2}x')
\phi(y-\frac{1}{2}x')k(x,u,x',u')e^{-i(ux'-u'x+u'y)}dxdx'du'
\end{equation}
and the choice of the kernel $k(x,u,x',u')$ selects one particular
function of the Cohen class. The Wigner function, for instance
arises for $k(x,u,x',u')=1$, whereas the ambiguity function is
obtained for $k(x,u,x',u')2\pi\delta(x-x')\delta(u-u')$.

\section{Lesser known  distribution function: the Kirkwood-Rihacek quasidistribution function}
Now we turn our attention to a lesser known distribution, the
Kirkwood-Rihaczek function, that may be written using the notation
above as \cite{Lee}
\begin{equation} K\left( \beta \right)
=\int d^{2}\alpha e^{\beta \alpha ^{\ast }-\beta ^{\ast }\alpha
}e^{\frac{\alpha ^{2}-\alpha ^{\ast 2}}{4}}C\left( \alpha \right)
\label{kirk} .
\end{equation}
This equation has been obtained from an equation similar to
equations (\ref{wigcar}), (\ref{anti}) and (\ref{normal}), i.e.
taking the double Fourier transform
\begin{equation} K\left( q,p \right)
=\int du dv e^{-iup} e^{ivq} Tr\{\rho e^{iv\hat{q}}e^{iu\hat{p}}\}
\end{equation}
and taking the trace as in equation (\ref{trace}) in the appendix.

 We will now do an analysis
similar to the one done in subsection $2.D.$ We relate the
Kirkwood-Rihaczek function to the Wigner function by using
(\ref{kirk}), via the following exponential of derivatives
\begin{equation}
K\left(
\beta \right)=e^{-\frac{1}{4}\frac{\partial ^{2}}{\partial ^{2}\beta }}e^{\frac{1}{4}\frac{%
\partial ^{2}}{\partial ^{2}\beta ^{\ast }}}W\left( \beta \right).
\end{equation}
We now use the non-integral expression for the Wigner function
\cite{series}
\begin{equation}
W\left( \beta \right) =Tr\left[ \left( -1\right) ^{\widehat{n}}\widehat{D}%
^{\dagger }\left( \beta \right) \widehat{\rho }\widehat{D}\left(
\beta \right) \right],
\end{equation}
with $\widehat{D}\left( \beta \right)=e^{\beta
a^{\dagger}-\beta^*a}$, the so-called Glauber displacement
operator. We cast the above equation  into the form
\begin{equation}
W\left( \beta \right) =Tr\left[ \left( -1\right)
^{\widehat{n}}\widehat{\rho }\widehat{D}\left( 2\beta \right)
\right],
\end{equation}
where we have used the trace property $Tr(AB)=Tr(BA)$ and the
following identities$\left( -1\right)
^{\widehat{n}}\widehat{D}^{\dagger }\left( \beta \right)
=\widehat{D}\left( \beta \right) \left( -1\right) ^{\widehat{n}}$.

Now we use the factorized form of the Glauber displacement
operator \cite{Louisel} $ \widehat{D}\left( 2\beta \right)
=e^{-2\left\vert \beta \right\vert ^{2}}e^{2\beta
\widehat{a}^{\dagger }}e^{-2\beta ^{\ast }\widehat{a}}$ to obtain
\begin{equation}
W\left( \beta \right) =Tr\left[ \left( -1\right)
^{\widehat{n}}\widehat{\rho }e^{-2\left\vert \beta \right\vert
^{2}}e^{2\beta \widehat{a}^{\dagger }}e^{-2\beta ^{\ast
}\widehat{a}}\right].
\end{equation}
Therefore, we have that the Kirkwood-Rihaczek function may be
written as

\begin{eqnarray}
\nonumber
K\left( \beta ,\beta ^{\ast }\right) &=&e^{-\frac{1}{4}\frac{\partial ^{2}}{%
\partial ^{2}\beta }}e^{\frac{1}{4}\frac{\partial ^{2}}{\partial ^{2}\beta
^{\ast }}}W\left( \beta ,\beta ^{\ast }\right) \\
&=&Tr\left[ \left( -1\right) ^{\widehat{n}}\widehat{\rho }e^{-\frac{1}{4}\frac{%
\partial ^{2}}{\partial ^{2}\beta }}e^{\frac{1}{4}\frac{\partial ^{2}}{%
\partial ^{2}\beta ^{\ast }}}\widehat{D}\left( 2\beta \right)
\right].
\end{eqnarray}
The calculation of the exponential of derivatives of the Glauber
operator will be  tedious but straightforward. We will just write
the main steps to obtain the final form, for instance, it is not
difficult to show that
\begin{equation}
e^{\frac{1}{4}\frac{\partial ^{2}}{\partial ^{2}\beta ^{\ast }}}\widehat{D}%
\left( 2\beta \right) =e^{-\beta ^{2}}e^{2\beta \left(
\widehat{a}^{\dagger
}+\widehat{a}-\beta ^{\ast }\right) }e^{\widehat{a}^{2}}e^{-2\beta ^{\ast }%
\widehat{a}}.
\end{equation}
By using that \cite{Arfken}
\begin{equation}
e^{-t^{2}+2tx}=\sum_{k=0}^{\infty }H_{k}\left( x\right)
\frac{t^{k}}{k!}
\end{equation}%
we can express the above equation as
\begin{equation}
e^{\frac{1}{4}\frac{\partial ^{2}}{\partial ^{2}\beta ^{\ast }}}\widehat{D}%
\left( 2\beta \right) =\sum_{k=0}^{\infty }H_{k}\left(
\widehat{a}^{\dagger
}+\widehat{a}-\beta ^{\ast }\right) \frac{\beta ^{k}}{k!}e^{\widehat{a}%
^{2}}e^{-2\beta ^{\ast }\widehat{a}},
\end{equation}%
with $H_k(x)$ the Hermite polynomials. From the above equation, is
easy to obtain
\begin{equation}
\frac{\partial ^{2n}}{\partial \beta ^{2n}}\sum_{k=0}^{\infty
}H_{k}\left( x\right) \frac{\beta ^{k}}{k!}=\sum_{k=0}^{\infty
}H_{k+2n}\left( x\right) \frac{\beta ^{k}}{k!}
\end{equation}
and therefore
\begin{equation}
e^{-\frac{1}{4}\frac{\partial ^{2}}{\partial ^{2}\beta }}e^{\frac{1}{4}\frac{%
\partial ^{2}}{\partial ^{2}\beta ^{\ast }}}\widehat{D}\left( 2\beta \right)
=\sum_{n=0}^{\infty }\sum_{k=0}^{\infty }\frac{\left(
-\frac{1}{4}\right) ^{n}}{n!}H_{k+2n}\left( \widehat{a}^{\dagger
}+\widehat{a}-\beta ^{\ast }\right) \frac{\left( \beta \right)
^{k}}{k!}e^{\widehat{a}^{2}}e^{-2\beta ^{\ast }\widehat{a}}.
\end{equation}
Now we use the integral form of the Hermite polynomials
\cite{Arfken}
\begin{equation}
H_{p}\left( x\right) =\frac{2^{p}}{\sqrt{\pi
}}\int\limits_{-\infty }^{\infty }\left( x+it\right)
^{p}e^{-t^{2}}dt
\end{equation}
to obtain
\begin{equation}
K\left( \beta ,\beta ^{\ast }\right) =\frac{e^{-\beta ^{\ast
2}}e^{-2\beta \beta ^{\ast }}}{\sqrt{\pi }}\int\limits_{-\infty
}^{\infty }dx\int\limits_{-\infty }^{\infty }dte^{\left(
-2\sqrt{2}x+2\beta ^{\ast }+2\beta \right)
it}e^{-2x^{2}}e^{2\sqrt{2}x\left( \beta ^{\ast }+\beta
\right) }\left\langle x\right\vert e^{\widehat{a}^{2}}e^{-2\beta ^{\ast }%
\widehat{a}}\left( -1\right) ^{\widehat{n}}\widehat{\rho
}\left\vert x\right\rangle
\end{equation}
by using
\begin{equation}
\int\limits_{-\infty }^{\infty }e^{-ikt}dt=2\pi \delta \left(
k\right)
\end{equation}
and taking $k=2\sqrt{2}x-2\beta ^{\ast }-2\beta $ we have
\begin{eqnarray}
K\left( \beta ,\beta ^{\ast }\right) &=&2\sqrt{\pi }e^{-\beta
^{\ast 2}}e^{-2\beta \beta ^{\ast }}\int\limits_{-\infty }^{\infty
}dx\delta
\left( 2\sqrt{2}x-2\beta ^{\ast }-2\beta \right) e^{-2x^{2}}e^{2\sqrt{2}%
x\left( \beta ^{\ast }+\beta \right) } \\
&&\times \left\langle x\right\vert e^{\widehat{a}^{2}}e^{-2\beta ^{\ast }%
\widehat{a}}\left( -1\right) ^{\widehat{n}}\widehat{\rho
}\left\vert x\right\rangle .
\end{eqnarray}
Making use of the identity $\delta \left( \alpha x\right) =\frac{\delta \left( x\right) }{%
\left\vert \alpha \right\vert }$ we finally obtain
\begin{equation}
K\left( \beta ,\beta ^{\ast }\right) =e^{-2\beta ^{\ast
2}}e^{-2\beta \beta
^{\ast }}\sqrt{\frac{\pi }{2}}e^{2\left( \frac{\beta ^{\ast }+\beta }{\sqrt{2%
}}\right) ^{2}}\left\langle \frac{\beta ^{\ast }+\beta }{\sqrt{2}}%
\right\vert e^{\left( \widehat{a}-\beta ^{\ast }\right)
^{2}}\left( -1\right) ^{\widehat{n}}\widehat{\rho }\left\vert
\frac{\beta ^{\ast }+\beta }{\sqrt{2}}\right\rangle
\end{equation}%
or
\begin{equation}
K\left( \beta ,\beta ^{\ast }\right) =\sqrt{\frac{\pi
}{2}}e^{\beta ^{2}-\beta ^{\ast 2}}\left\langle X
\right\vert D^{\dagger }\left( -\beta ^{\ast }\right) e^{\widehat{a}%
^{2}}D\left( -\beta ^{\ast }\right) \left( -1\right) ^{\widehat{n}}\widehat{%
\rho }\left\vert X\right\rangle
\end{equation}%
with $X=\frac{\beta ^{\ast }+\beta }{\sqrt{2}}$.

In this form, we have succeeded in obtaining the Kirkwood-Rihaczek
function as an expectation value, in terms of position
eigenstates, just as the $Q$-function, in terms of coherent states
[see equation (5)], the Wigner and Glauber-Sudarshan functions in
term of number states \cite{series}.
\section{Conclusions}
We have shown that some distribution functions may be related
through a method that allows the construction of some
quasi-probability  functions such as the Wigner, Glauber-Sudarshan
and Husimi functions \cite{Puri}. This method consists of
obtaining the distribution functions from a double Fourier
transform of an averaged exponential operator. If we use the
exponential operator in terms of creation and annihilation
operators we construct the already mentioned distribution
functions. Via this method, but leaving the exponential operator
in terms of position and momentum, and ordering (factorizing) the
exponential in a convenient way, another function lesser used in
classical optics, namely, Kirkwood-Rihacek's distribution function
may be obtained. This function was recently introduced in quantum
mechanics by Praxmeyer and W\'odkiewicz
\cite{Praxmayer1,Praxmayer2} to have a phase representation of the
Hydrogen atom. The connection between Glauber-Sudarshan and Husimi
functions and functions of the Cohen class has been given, i.e.
the adequate kernels. Finally, the Kirkwood-Rihacek has been given
in term of an expectation value, in terms of position eigenstates,
just as other distribution functions may also be given in terms of
(sums of) expectation values. This may be of interest because it
has been already exploited the fact that these forms allow
reconstruction of quasiprobability distribution functions
\cite{Moya2,Moya3}.

\section{Appendix} \label{app} In Dirac notation, we denote  functions
"$f$" by means of "kets" $|f\rangle$. For instance an
eigenfunction of the harmonic oscillator \cite{Arfken}
\begin{equation}\label{eigfun}
\psi_n(x)=\frac{\pi^{-1/4}}{\sqrt{2^{n}n!}}\mathrm{e}^{-x^2/2}H_n(x).
\end{equation}
is represented by the ket $|n\rangle$, with $n=0,1,2, ...$ In
quantum mechanics, these states are called number or Fock states
(see for instance \cite{Schleich}). Any function can be expanded
in terms of eigenfunctions of the harmonic oscillator
\begin{equation}
f(x)=\sum_{n=0}^{\infty}c_n \psi_n(x)
\end{equation}
where
\begin{equation}
c_n =\int_{-\infty}^{\infty}f(x)\psi_n(x) dx
\end{equation}
and in the same way any ket may be expanded in terms of
$|n\rangle$'s
\begin{equation}
|f\rangle=\sum_{n=0}^{\infty}c_n |n\rangle
\end{equation}
where the orthonormalization relation
\begin{equation}
\langle m |n\rangle = \int_{-\infty}^{\infty}\psi^*_m(x) \psi_n(x)
dx = \delta_{nm}
\end{equation}
has been used. The quantity $\langle m |$ is a so-called "bra".

The basis set of kets $|n\rangle$ is a discrete one. However,
there are also continuous basis. We can form one continuous basis
for example with the function $e^{ipq}/\sqrt{2\pi}$ and the
corresponding ket $|p\rangle$. First note that
\begin{equation}
\langle p |p'\rangle = \frac{1}{2\pi}\int_{-\infty}^{\infty}
e^{-i(p-p')q}dx = \delta(p-p'),
\end{equation}
so that
\begin{equation}
e^{ip'q}/\sqrt{2\pi}= \int_{-\infty}^{\infty}
\delta(p-p')e^{ipq}/\sqrt{2\pi}dp,
\end{equation}
or, in bra-ket notation we have
\begin{equation}
 |p'\rangle = \int_{-\infty}^{\infty}
 \delta(p-p')|p\rangle dp =\int_{-\infty}^{\infty}
 \langle p |p'\rangle |p\rangle dp,
\end{equation}
rearranging terms we have
\begin{equation}
 |p'\rangle  =\left(\int_{-\infty}^{\infty}
 |p\rangle\langle p | dp\right) |p'\rangle = {\bf 1} |p'\rangle ,
\end{equation} i.e. we have the completeness relation
\begin{equation}
\int_{-\infty}^{\infty}
 |p\rangle\langle p | dp= {\bf 1}  .
\end{equation}
Finally note that the function $e^{ipq}/\sqrt{2\pi}$ is an
eigenfunction of the operator $-i\frac{d}{dq}$ with eigenvalue
$p$.
 For position, an "eigenket"  of $\hat{q}$ is
\begin{equation}
\hat{q} |q\rangle = q |q\rangle
\end{equation}
and an "eigenbra"
\begin{equation}
 \langle q'|q' =  \langle q'| \hat{q}.
\end{equation}
We therefore find
\begin{equation}
\langle q'|q  \rangle (q'-q)=   0,
\end{equation}
that has as solution \cite{Schleich}
\begin{equation}
\langle q'|q  \rangle = \delta(q'-q).
\end{equation}
We then can express the completeness relation
\begin{equation}
{\bf 1}= \int_{-\infty}^{\infty}|q  \rangle\langle q| dq   ,
\end{equation}
such that
\begin{equation}
|\Psi\rangle={\bf 1}|\Psi\rangle=\int_{-\infty}^{\infty}|q
\rangle\langle q||\Psi\rangle dq =
\int_{-\infty}^{\infty}\Psi(q)|q\rangle
 dq
\end{equation}
where $\Psi(q)=\langle q||\Psi\rangle=\langle q|\Psi\rangle$. the
density matrix $\rho$ is defined simply as the ket-bra operator
$\rho=|\Psi\rangle \langle \Psi |$.

The completeness relation serve us among other things to calculate
averages, for instance
\begin{equation}
\langle \Psi| \hat{A}|\Psi\rangle= \langle \Psi| \hat{A}{\bf
1}|\Psi\rangle =\langle \Psi| \hat{A}\int_{-\infty}^{\infty}|q
\rangle\langle q| dq|\Psi\rangle = \int_{-\infty}^{\infty}\langle
\Psi|\hat{A}|q \rangle\langle q| \Psi\rangle dq
\end{equation}
or finally,
\begin{equation}
\langle \Psi| \hat{A}|\Psi\rangle= \int_{-\infty}^{\infty}\langle
q| \Psi\rangle\langle \Psi|\hat{A}|q \rangle dq. \label{trace}
\end{equation}
Note that in the above equation we are simply adding "diagonal"
elements, i.e. we have the trace of the the operator
$|\Psi\rangle\langle \Psi|\hat{A}$. As the trace is independent of
the basis, we can have it in terms of the discrete basis
$|n\rangle$
\begin{equation}
\langle \Psi| \hat{A}|\Psi\rangle= \sum_{n=0}^{\infty}\langle n|
\Psi\rangle\langle \Psi|\hat{A}|n \rangle\equiv Tr\{\rho \hat{A}\}
.
\end{equation}

\end{document}